# Simultaneous emergence of superconductivity, inter-pocket scattering and nematic fluctuation in potassium-coated FeSe superconductor

Z. R. Ye[1, †], C. F. Zhang[2, 3, †], H. L. Ning[1], W. Li[2, 3], L. Chen[1], T. Jia[2, 3], M. Hashimoto[4], D. H. Lu[4], Z.-X. Shen[2, 3], and Y. Zhang[1, 5, *]

[1]*International Center for Quantum Materials, School of Physics, Peking University, Beijing 100871, China*

[2]*Stanford Institute for Materials and Energy Sciences, SLAC National Accelerator Laboratory, 2575 Sand Hill Road, Menlo Park, California 94025, USA*

[3]*Geballe Laboratory for Advanced Materials, Departments of Physics and Applied Physics, Stanford University, Stanford, California 94305, USA*

[4]*Stanford Synchrotron Radiation Lightsource, SLAC National Accelerator Laboratory, 2575 Sand Hill Road, Menlo Park, California 94025, USA*

[5]*Collaborative Innovation Center of Quantum Matter, Beijing 100871, China*

† These authors contribute equally to this work.

* To whom correspondence should be addressed: yzhang85@pku.edu.cn

**Superconductivity originates from pairing of electrons. Pairing channel on Fermi surface and pairing glue are thus two pivotal issues for understanding a superconductor. Recently, high-temperature superconductivity over 40 K was found in electron-doped FeSe superconductors including $K_xFe_{2-y}Se_2$, $Li_{0.8}Fe_{0.2}OHFeSe$, and 1 monolayer FeSe thin film. However, their pairing mechanism remains controversial. Here, we studied the systematic evolution of electronic structure in potassium-coated FeSe single crystal. The doping level is controlled precisely by *in situ* evaporating potassium onto the sample surface. We found that the superconductivity emerges when the inter-pocket scattering between two electron pockets is turned on by a Lifshitz transition of Fermi surface. The nematic order suppresses remarkably at the same doping and strong nematic fluctuation remains in a wide doping range of the phase diagram. Our results suggest an underlying correlation among superconductivity, inter-pocket scattering, and nematic fluctuation in electron-doped FeSe superconductors.**

Most iron-pnictide superconductors share the same Fermi surface topology. The Fermi surface consists of both hole and electron pockets at the center and corner of the Brillouin zone (BZ) and the inter-pocket scattering between the hole and electron pockets is believed to be critical for the high-$T_c$ superconductivity[1-3]. However, in electron-doped FeSe superconductors, the heavy electron doping pushes the hole bands well below the Fermi energy ($E_F$) and the Fermi surface consists only of electron pockets at the BZ corner[4-7]. Surprisingly, high-$T_c$ superconductivity emerges on such a Fermi surface. The $T_c$ is over 30 K in $K_xFe_{2-y}Se_2$ and $Li_{0.8}Fe_{0.2}OHFeSe$[8,9], and possible superconductivity over 65 K was observed in 1 monolayer (ML) FeSe thin film[6, 7, 10]. The discovery of high-$T_c$ superconductivity in electron-doped FeSe superconductors challenges our previous understanding on iron-based superconductor. It is still not clear how the pairs of electron scatter on the Fermi surface and what fluctuation mediates the superconductivity. The answer to these questions immediately connects to the pairing mechanism of high-$T_c$ superconductivity in electron-doped FeSe superconductors and needs to be settled experimentally.

One effective way to understand a superconductor is to characterize its phase diagram. However, the doping level is difficult to control in electron-doped FeSe superconductors. Discrete phases are found in $K_xFe_{2-y}Se_2$ due to its intrinsic phase separation[11]. The electron doping of 1ML FeSe is believed to originate from the oxygen vacancy in $SrTiO_3$ substrate and its fine control has not been achieved so far[7]. On the other hand, efforts have been made to dope electrons into iron-pnictide compounds in order to remove the central hole pockets and reproduce the Fermi surface topology that is similar to electron-doped FeSe superconductors. However, superconductivity has not been realized in most electron-doped iron-pnictide compounds[12, 13]. Candidates such as $LaO_{1-x}H_xFeAs$ and $Ca_{1-x}La_xFe_2As_2$ are rarely studied due to the lack of high-quality single crystal and possible low superconducting volume[14, 15].

Recently, potassium evaporation has been found to be an effective method that can dope electrons into a sample[16]. Utilizing such a method, electron doping is realized in FeSe thin film and the emergence of superconductivity associated with the suppression of nematicity was observed[16-18]. Here, we performed angle-resolved photoemission spectroscopy (ARPES) measurements on potassium-coated FeSe single crystals. By *in situ* evaporating a small amount of potassium onto the sample surface step by step, we successfully doped electrons into FeSe single crystal with unprecedented precision (See details in Methods section). We found that the Fermi surface goes through two Lifshitz transitions with electron doping. The hole pocket first vanishes at BZ center which triggers the first Lifshitz transition. Then, by further electron doping, one electron band sinks below $E_F$ at the BZ corner resulting in the second Lifshitz transition. Intriguingly, the superconductivity emerges at the second Lifshitz transition when the inter-pocket scattering between the electron pockets is turned on Meanwhile, the nematic order suppresses remarkably and the residual nematic order is observed penetrating deeply into the superconducting dome. Our result provides crucial clues for understanding the high-Tc superconductivity in electron-doped FeSe superconductors. The pairing of electron occurs through inter-pocket scattering between two electron pockets and the nematic fluctuation plays an important role in mediating the superconducting pairing.

**Result**

**Evolution of band structure with electron doping.** The doping process is divided into 15 fine steps (D0-D14). The doping dependence of the second derivative photoemission spectra is shown in Fig.1. The sharpness of the spectra indicates the high quality of our sample surface during the evaporation. At zero doping (D0), two hole-like bands ($\beta_1$ and $\beta_2$) are observed around the BZ corner, the ($\pi$, $\pi$) point. They correspond to the split $d_{xz}$ and $d_{yz}$ bands stemming from the band reconstruction in the nematic state[19-21]. Around the BZ zone center, the (0, 0) point, three hole-like bands (α, β, and γ) are observed. They are constructed by the $d_{xz}$, $d_{yz}$, and $d_{xy}$ orbitals respectively[21]. Similar to the $β_1/β_2$ bands at (π, π), the β band splits near $E_F$ due to the presence of nematic order. The γ band disperses across the α and β bands. Hybridization gap opens at the band crossing between the β and γ bands.

When doping electrons into the system, bands shift downwards to higher binding energy as expected (Fig. 1). However, one set of bands remains unchanged and resembles the bulk FeSe bands even in heavily doped sample (Fig. 1h). One possible explanation is the potassium inhomogeneity on the sample surface. However, the doping inhomogeneity usually generates multiple sets of bands with different doping levels, which is inconsistent with the observation of only two sets of bands. On the other hand, because the probing depth of ARPES can reach several layers below the sample surface, our results suggest that potassium evaporation only dose the topmost layer, while the layers beneath remain undoped (Fig. 1i). The invariant bands are thus attributed to the undoped bulk state.

Figure 2 illustrates how bands and Fermi surface evolve with electron doping in potassium-coated FeSe single crystal. For the hole bands at the (0, 0) point, the band tops of α and β move to higher binding energy as characterized by the energy distribution curves (EDCs) taken at the (0, 0) point (Fig. 2a and 2b). We note that, all the data were taken with 23 eV photons. The corresponding $k_z$ is around the Z point of BZ[21], where the band tops of the hole bands are the highest. Therefore, when the β band sinks below $E_F$ at D3, the entire hole pocket disappears and the first Lifshitz transition of Fermi surface occurs. Comparing with the α and β bands, the γ band shows weak photoemission intensity and therefore cannot be tracked directly from the spectra. However, we can determine the energy position of the γ band through the hybridization gap between the γ and β bands. As shown in Figs. 1b, 1e, and 1h, the energy position of the hybridization exhibits no observable change implicating that the γ band remains unchanged during the doping process (Fig. 2c).

We then turn to the bands around the (π, π) point. The energy splitting between the $β_1$ and $β_2$ bands reflects the strength of nematic order. As shown in Fig. 2d and 2e, the $β_1$ and $β_2$ bands shift to higher binding energy and meanwhile their energy splitting decreases. A kink is observed for the band shift of $β_2$, which indicates a remarkable suppression of nematic order at D5. Besides the $β_1$ and $β_2$ bands, three electron bands are observed during the doping process (Fig. 1 and 2f). The δ and ζ bands construct the highly anisotropic electron pocket at the BZ corner (Fig. 2g) and are originated from the $d_{yz}$ and $d_{xy}$ orbitals respectively[20]. As shown in Figs. 1b, 1e and 1h, when doping electrons, the δ band shifts downward and eventually merges with the ζ band. As a result, the ζ/δ electron pocket becomes isotropic and circular-like in heavily doped samples (Fig. 2g). Besides the δ and ζ bands, the η band is above $E_F$ and thus cannot be observed in the lightly doped samples. With electron doping, the η band sinks below $E_F$ at D5 (Figs. 1, 2e and 2d), triggering the second Lifshitz transition of Fermi surface. The η band constructs a circular-like electron pocket and is originated from the mixing of $d_{xz}$ and $d_{xy}$ orbitals[20]. The η electron pocket emerges at the second Lifshitz transition and overlaps with the δ/ζ electron pockets at the BZ corner (Fig. 2g). The band structure of heavily electron-doped FeSe single crystal resembles that observed in $K_xFe_{2-y}Se_2$, $Li_{0.8}Fe_{0.2}OHFeSe$, and 1 monolayer FeSe thin film[7-10]. The band bottom of all electron bands degenerates at around 50meV below $E_F$, which implicates that the ζ band bottom does not shift during the electron doping process. The invariability of γ and ζ suggests that the occupation of $d_{xy}$ orbital is barely affected by the electron doping in contrast to the $d_{xz}$ and $d_{yz}$ orbitals. Such orbital-selectivity may originate from the orbital-selective correlation effect in FeSe[22, 23]. In contrast to the strongly correlated $d_{xy}$ orbital, electrons prefer to fill in the more itinerate $d_{xz}$ and $d_{yz}$ orbitals.

**Evolution of superconductivity with electron doping.** After characterizing the band structure, it is then intriguing to study how superconductivity evolves during the doping process. The EDCs are taken at the Fermi crossing ($k_F$) of the ζ band along (0, 0) to (π, π) direction. To remove the effect of Fermi distribution function near $E_F$, we symmetrize the EDCs as shown in Figs. 3a and 3b. The superconducting gap opening is characterized by the emergence of an intensity dip at $E_F$ and two coherence peaks in the symmetrized EDCs. The gap magnitude is determined by the gap fitting using a phenomenology spectral function[24]. The superconducting gap emerges at D5, then increases to its maximum around 8 meV at D11, and finally decreases to around 7 meV at D14. The $T_c$ is characterized by the onset temperature of gap opening and the suppression of spectral weight at $E_F$ as shown in Figs. 3c-3h. The $T_c$ correlates with the gap magnitude showing dome-like behavior. Specifically, it is 18.5 K at D8, 25 K at D11, and 20 K at D14.

**Phase diagram.** Considering that most microscopic models of iron-based superconductors are constructed in one-iron BZ[1-3], we unfolded the Fermi surface into one-iron BZ and the transition of Fermi surface topology is more transparent. The disappearance of β hole pocket and the emergence of η electron pocket trigger two Lifshitz transitions of Fermi surface (Fig. 4a). For the type I Fermi surface, the inter-pocket scattering connects the hole and electron pockets through a $(\pi, 0)$ momentum transfer, while for the type III Fermi surface, the inter-pocket scattering occurs between the two electron pockets. Its scattering vector is $(\pi, \pi)$ in one-iron BZ, which corresponds to a small momentum transfer in two-iron BZ. In the middle, the inter-pocket scattering is strongly suppressed and only intra-pocket scattering presents.

Figure 4b shows the phase diagram of potassium-coated FeSe single crystal. We estimate the electron doping level by counting the Fermi surface volume. The nematic transition temperature ($T_{nem}$) is calculated from the energy splitting between the $\beta_1$ and $\beta_2$ bands ($\Delta_{nem}$). The $\Delta_{nem}/k_B T_{nem}$ is 6.22 as determined by the $\Delta_{nem}$ and $T_{nem}$ in undoped FeSe single crystal. The doping dependence of $\Delta_{nem}/k_B T_{nem}$ is ignored. Similarly, the $T_c$ is calculated from the superconducting gap magnitude ($\Delta_{sc}$) and the $2\Delta_{sc}/k_B T_c$ is 7.74 as determined from the data in Fig.3. When doping electrons, a dome-like superconducting phase ($SC_{e-e}$) emerges at the second Lifshitz transition. Its optimal doping is around 0.1 electrons per iron and the maximal $T_c$ is around 25K. Note that, the gap is not observed in lightly doped samples due to the limitations of the energy resolution and the lowest achievable sample temperature in our experiment. The $SC_{h-e}$ dome is illustrated according to the recent scanning tunneling microscopy (STM) study on FeSe thin film grown on graphitized SiC whose properties resemble the bulk FeSe[26].

The $T_{nem}$ decreases remarkably at the second Lifshitz transition but does not reach zero (Fig 4b). The residual nematic order shows a tail-like behavior penetrating deeply into the $SC_{e-e}$ dome. When the nematic fluctuation is strong in a system, an external uniaxial pressure can stabilize the fluctuation resulting in an emergent nematic order. This has been observed near the nematic transition along the temperature axis in detwinned iron-pnictides[19]. Here in potassium-coated FeSe single crystal, the doping dependence of $T_{nem}$ can be explained by a smearing of nematic transition along the doping axis due to the presence of uniaxial pressure (inset of Fig. 4b). As discussed in Fig. 1i, the bottom layers are undoped and therefore remain C2 symmetric. The symmetry breaking of the bottom layers applies a uniaxial pressure on the topmost layer, which smears the sharp suppression of nematic order at around 0.03 electron doping resulting in a tail-like residual nematic order. Such emergent nematic order can be observed until 0.1 electron doping, which implicates that strong nematic fluctuation presents in a wide doping range in the topmost layer of potassium-coated FeSe single crystal.

**Discussion**

The Lifshitz transition of Fermi surface, the emergence of superconductivity and the suppression of nematic order all occurs at the same doping level, which suggests an underlying correlation among them. The $SC_{e-e}$ phase emerges when the inter-pocket scattering between two electron pockets is turned on by the second Lifshitz transition. This suggests that the inter-pocket scattering is crucial for the superconducting pairing in the $SC_{e-e}$ phase. Such inter-pocket scattering can be mediated by spin fluctuation as proposed in both weak- and strong-coupling theories[27, 28]. However, the long-range magnetic order has not been observed in FeSe. It is still not clear how strong the spin fluctuation is and whether the absence of magnetic order in FeSe is due to quantum fluctuation or not[29]. Alternatively, the nematic fluctuation may mediate the inter-pocket scattering through a small momentum transfer in two-iron BZ. Our result suggests that strong nematic fluctuation presents in potassium-coated FeSe single crystal when the nematic order is suppressed by electron doping. Intriguingly, when the nematic order at zero doping is enhanced from FeSe single crystal ($T_{nem}$ = 90K) to FeSe thin film ($T_{nem}$ = 160~180K), the optimal doping $T_c$ increases from 25 K to around 44~48 K and the superconducting gap increases from 8 meV to 12~14 meV[19-21]. All these suggest that the nematic order is most likely the parent phase for electron-doped FeSe superconductors. It is expected that $T_c$ would be further increased if the nematic fluctuation can be further enhanced.

Due to its distinctive Fermi surface topologies, the potassium-coated FeSe single crystal serves as a model system for further experimental and theoretical studies. The $SC_{h-e}$ dome ends at the first Lifshitz transition where the central hole pocket vanishes implying that the inter-pocket scattering between hole and electron pockets is critical for the $SC_{h-e}$ phase. Furthermore, the $SC_{h-e}$ phase coexists with the nematic order and its superconducting dome is well beneath the nematic transition. This suggests that, unlike the $SC_{e-e}$ phase, the nematic fluctuation is absent above $T_c$ for the $SC_{h-e}$ phase due to the formation of long-range nematic order. Therefore, our result suggests possible different pairing mechanisms for the $SC_{e-e}$ and $SC_{h-e}$ phases in potassium-coated FeSe single crystal. The two superconducting phases ($SC_{e-e}$ and $SC_{h-e}$) represent two distinct superconducting pairing channels in iron-based superconductors. It is intriguing to further study how they interplay in iron-pnictide superconductors when both hole-electron and electron-electron inter-pocket scatterings exist on the Fermi surface.

**Methods**

High quality FeSe single crystal was synthesized using vapor transport method[25]. The superconducting transition temperature is 8.4 K and the nematic transition temperature is 90K as characterized by magnetic susceptibility measurements. The potassium was evaporated onto the cleaved sample surface *in-situ*. The pressure is less than $5\times10^{-10}$ torr during the evaporation. ARPES measurements were performed at the Stanford Synchrotron Radiation Lightsource Beamline 5-4. All the data were taken with 23 eV photons. The temperature was maintained at 6 K for the low temperature measurement. The overall energy resolution was 5 meV, and the angular resolution was less than 0.3 degree. All the samples were measured in ultrahigh vacuum with a base pressure better than $3\times10^{-11}$ torr.

**Acknowledgements**

We gratefully thank F. Wang and D.-H. Lee for stimulating discussions. This work is supported by National Science Foundation of China (NSFC) under the grant No. 91421107. Stanford Synchrotron Radiation Lightsource is operated by the Office of Basic Energy Sciences, U.S. Department of Energy.

**Author Contributions**

Z. R. Y. and Y. Z. conceived the project. Z. R. Y., L. C. and Y. Z. synthesized the single crystals. C. F. Z., H. L. N., W. L., J. T. and Y. Z. took the ARPES measurements. Z. R. Y and

Y. Z. analyzed the data. M. H., D. H. L., and Z. X. S supported the ARPES measurements at Stanford Synchrotron Radiation Lightsource. Z. R. Y, H. L. N and Y. Z. wrote the paper with the input from all coauthors.

**Author Information**

The authors declare no competing financial interests. Correspondence and request for materials should be addressed to Y. Zhang (yzhang85@pku.edu.cn).

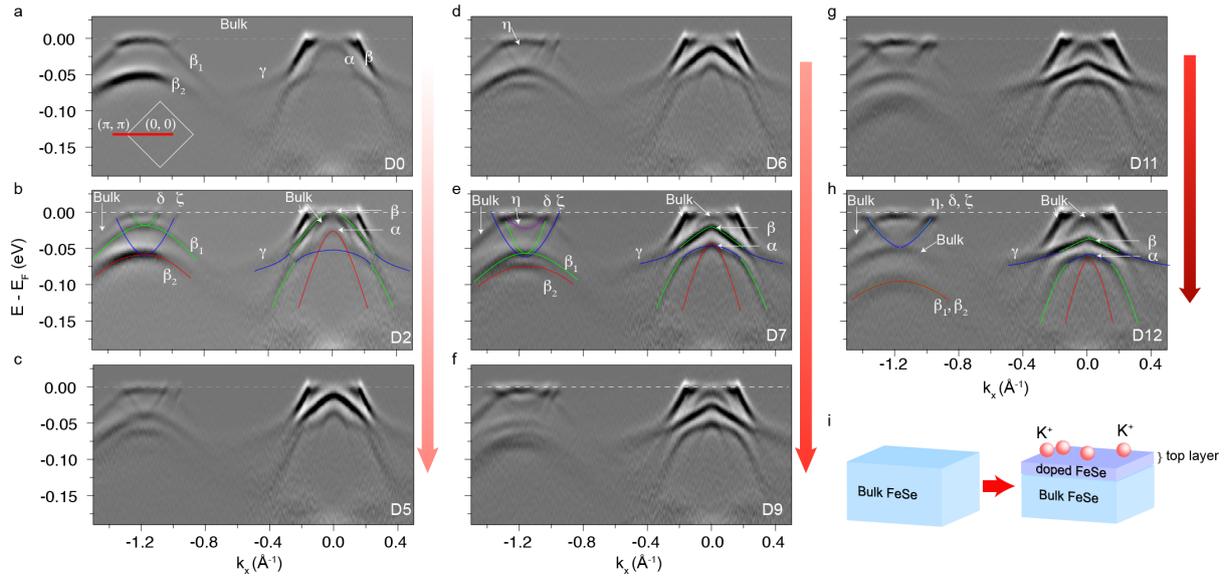

**Figure 1 | Evolution of the photoemission spectra in potassium-coated FeSe single crystal. a-h**, Doping dependence of the second derivative photoemission spectra taken along the (0, 0)-(π, π) high symmetry direction. The cut momentum is illustrated by the red solid line in the inset of panel a. We use D0-D14 to denote the doping sequence, where D0 represents the undoped sample. For the convenience to observe the evolution of the bands, we label out the bulk states (Bulk) and different bands (α, β, γ, δ, ζ, $β_1$, $β_2$, and η) in the representative panels. **i,** Schematic of potassium doping process. Only the top layer is dosed by the potassium ions while the bottom layers remain undoped.

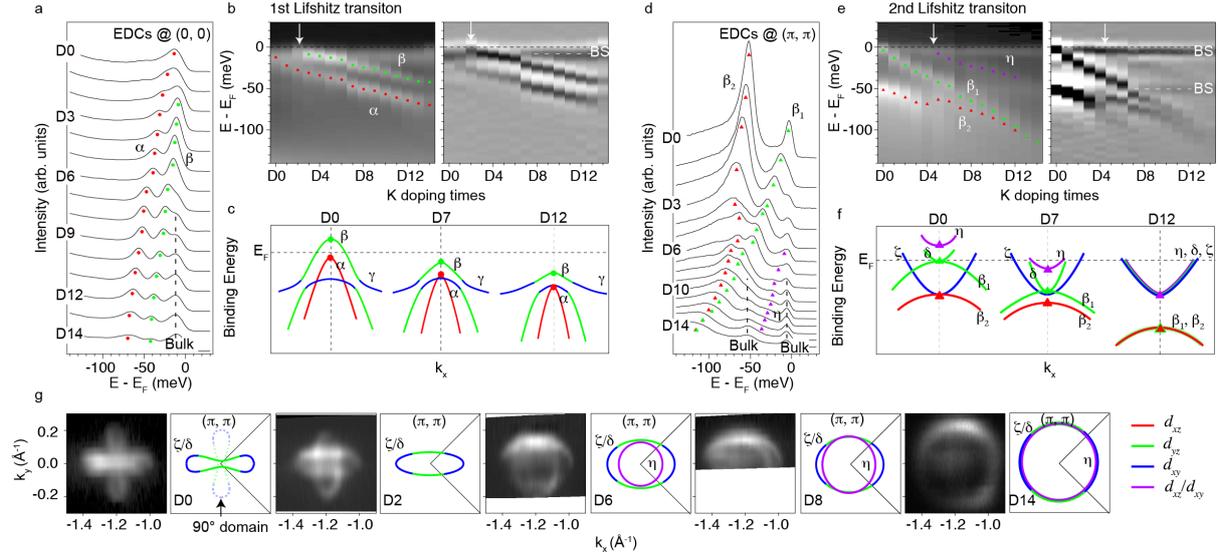

**Figure 2 | Evolution of the band structure and Fermi surface in potassium-coated FeSe single crystal. a,** Doping dependence of the energy distribution curves (EDCs) taken at the (0, 0) point. **b,** Merged image (left panel) and its second derivate image (right panel) of the EDCs in panel a. The peak positions of α and β are illustrated using red and green circles. The dashed line is the guide to eyes for following the bulk state. The white arrow illustrates the first Lifshitz transition. **c,** Illustration of the α, β, and γ bands at D0, D7, and D12. The color of bands reflects their orbital characters. **d-f** are the same as panel **a-c** but taken around the (π, π) point. **g,** Fermi surface mapping and the corresponding topology schematics taken around the (π, π) point at D0, D2, D6, D8, and D14. Note that, our spectra consist of signals from two perpendicular domains due to the sample twinning effect in the nematic state.

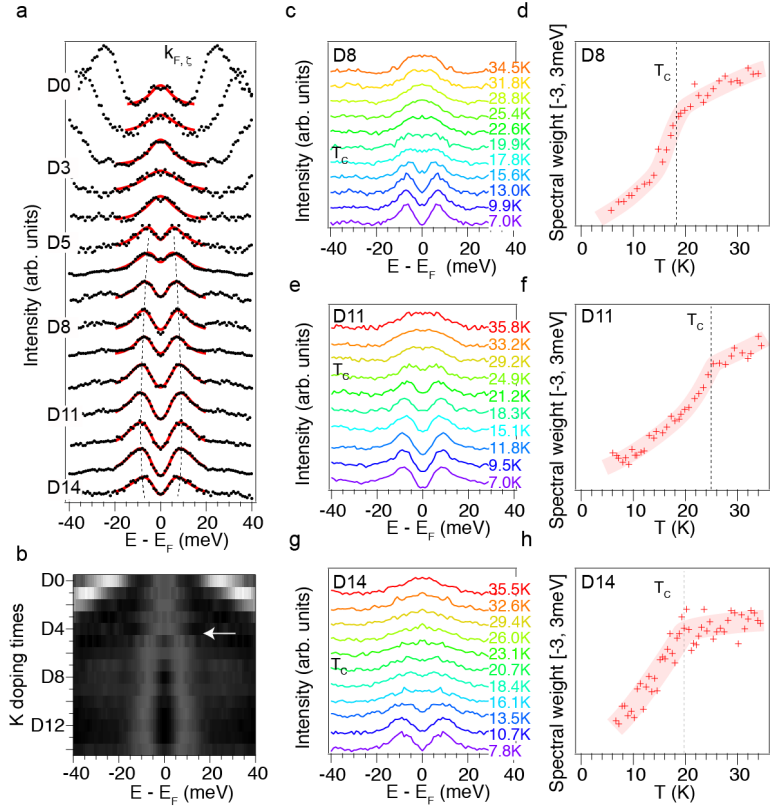

**Figure 3 | Evolution of superconductivity in potassium-coated FeSe single crystal. a,** Doping dependence of the symmetrized EDCs taken at the Fermi crossing (k$_F$) of the ζ band at 6 K. The gap size is obtained by fitting the symmetrized EDCs using a phenomenology function[24]. The fitting results are shown by the red solid lines. The dashed lines are the guides to eye of the superconducting peak positions. **b,** The merged image of the symmetrized EDCs in panel a. The white arrow indicates the emergence of superconducting gap. **c,** Temperature dependence of the symmetrized EDCs taken at the k$_F$ of the ζ band at D8. The T$_c$ is determined by the gap opening temperature as characterized by the dip emergence at E$_F$. **d,** Temperature dependence of the spectra weight integrated near E$_F$ within [-3, 3 meV] energy window. The T$_c$ is determined by the sharp suppression of spectral weight at E$_F$. **e, f, g,** and **h** are the same as **c** and **d**, but taken at D11 and D14.

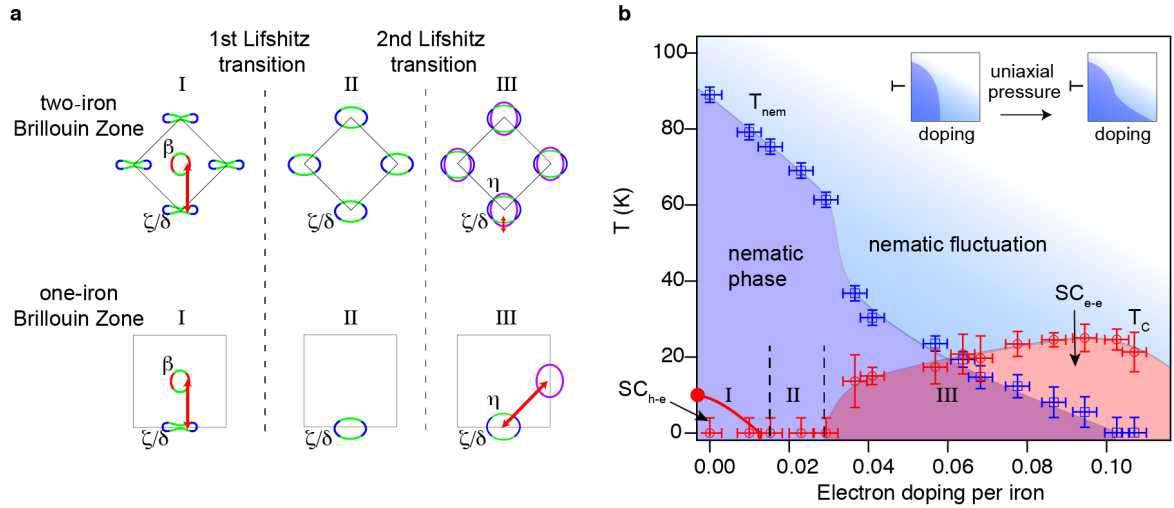

**Figure 4 | Phase diagram of potassium-coated FeSe single crystal. a**, the Lifshitz transitions of Fermi surface in two-iron (upper panel) and one-iron (lower panel) BZ. The red arrows denote the vectors of inter-pocket scattering. **b**, Phase diagram of potassium-coated FeSe single crystal. The black dashed lines represent the two Lifshitz transitions. The $T_c$ is around 8.4 K for bulk FeSe[25] and the $SC_{h-e}$ dome is illustrated by red solid line according to ref. 26.